\documentclass[10pt]{article} 
\usepackage{graphicx}

\def\rv{{\bf r}}
\def\vv{{\bf v}}

\def\pv{{\bf p}}

\def\Ev{{\bf E}}

\def\Bv{{\bf B}}
\def\Av{{\bf A}}

\def\xu{\hat{\bf x}}
\def\yu{\hat{\bf y}}

\def\bcl{b_{\rm cl}}

\def\BB{b_0}

\def\Airi{{\cal A}}

\def\lambdabar{\lambda\raise0.4ex\hbox{\kern-0.5em\hbox{--}}\ }
\def\lambdaC{\lambda\raise0.5ex\hbox{\kern-0.5em\hbox{--}}_{\rm C}}
\def\lambdabarc{\lambda\raise0.5ex\hbox{\kern-0.5em\hbox{--}}_{\rm c}}
\def\lesssim{{\lower0.5ex\hbox{$\stackrel{<}{\sim}$}}}
\def\gtrsim{{\lower0.5ex\hbox{$\stackrel{>}{\sim}$}}}

\begin{document}

{\fontsize{17.28}{24}\selectfont{\bf
\centerline{Analogy between free electron laser}
\centerline{and channeling by crystal planes}
}}
\vskip 0.6 true cm

{\fontsize{12}{16}\selectfont 
\centerline{Xavier ARTRU\footnote{e-mail: x.artru@ipnl.in2p3.fr}}
\centerline{Institut de Physique Nucl\'eaire de Lyon,} 
\centerline{Universit\'e Claude-Bernard \& IN2P3-CNRS,}
\centerline{69622 Villeurbanne, France}
}
\vskip 0.8 true cm

{\fontsize{11}{15}\selectfont \centerline{{\bf 
ABSTRACT
}}}

\medskip\noindent
The trapping of electrons in the ponderomotive potential wells, 
which governs a free electron laser or inverse free electron laser at high gain,
is analogous to the channeling of charged particles by atomic planes of a crystal.
A bent crystal is analogous to a period-tapered free electron laser.
This analogy is different from the well-known one between channeling and undulator radiations.

\medskip\noindent
{\bf keywords:}
free electron laser, inverse free electron laser, tapered undulator, channeling, bent crystal

\bigskip\noindent
{\fontsize{11}{15}\selectfont \centerline{{\bf 
1. INTRODUCTION
}}}

\bigskip\noindent
When a fast charged particle enters a crystal parallel to, or at small angle with, 
a family of atomic planes (see Fig.1), it is gently but firmly deflected by successive 
collisions with atoms belonging to a same plane \cite{channeling}. 
The particle is essentially subject to the so-called {\it continuous potential} $V(x)$, 
which is the electrostatic crystal potential averaged over the coordinates 
$y$ and $z$ parallel to the planes.
Particles trapped in a well of the continuous potential are said to be {\it channeled}. 
A similar phenomenon can happen in a free electron laser (FEL): 
the electron can be trapped in a well of the so-called $ponderomotive$ potential
which results from the superposition of the undulator magnetic field 
and the incoming radiation field. 
Channeling in crystal deflects the particle without changing its energy, 
whereas a FEL decelerates it without changing the transverse momentum, 
but otherwise, they have more than one analogous properties.
We will discuss some of the latter below in very simplified but parallel 
presentations of channeling and FEL. We think that such a presentation 
may help to understand the basic mechanism of the free electron laser
(for a more standard presentation, see, for instance \cite{Colson85}).

We must warn the reader that the analogy discussed in this paper must not be confused with the 
well-established similarity between {\it channeling radiation} and undulator radiation.

\bigskip\noindent
{\fontsize{11}{15}\selectfont{\bf 
\centerline{2. MOMENTUM EXCHANGE BETWEEN A CRYSTAL}
\centerline{AND A CHANNELED PARTICLE}
}}

\medskip\noindent
The continuous potential energy $U(x) = Ze V(x)$ 
is periodic and repulsive from the planes for positive particles, 
attractive to the planes for negative particles ($Ze$ is the particle charge).
\begin{figure}[ht]
\includegraphics*[scale=0.8,clip,bb=115 485 320 735]{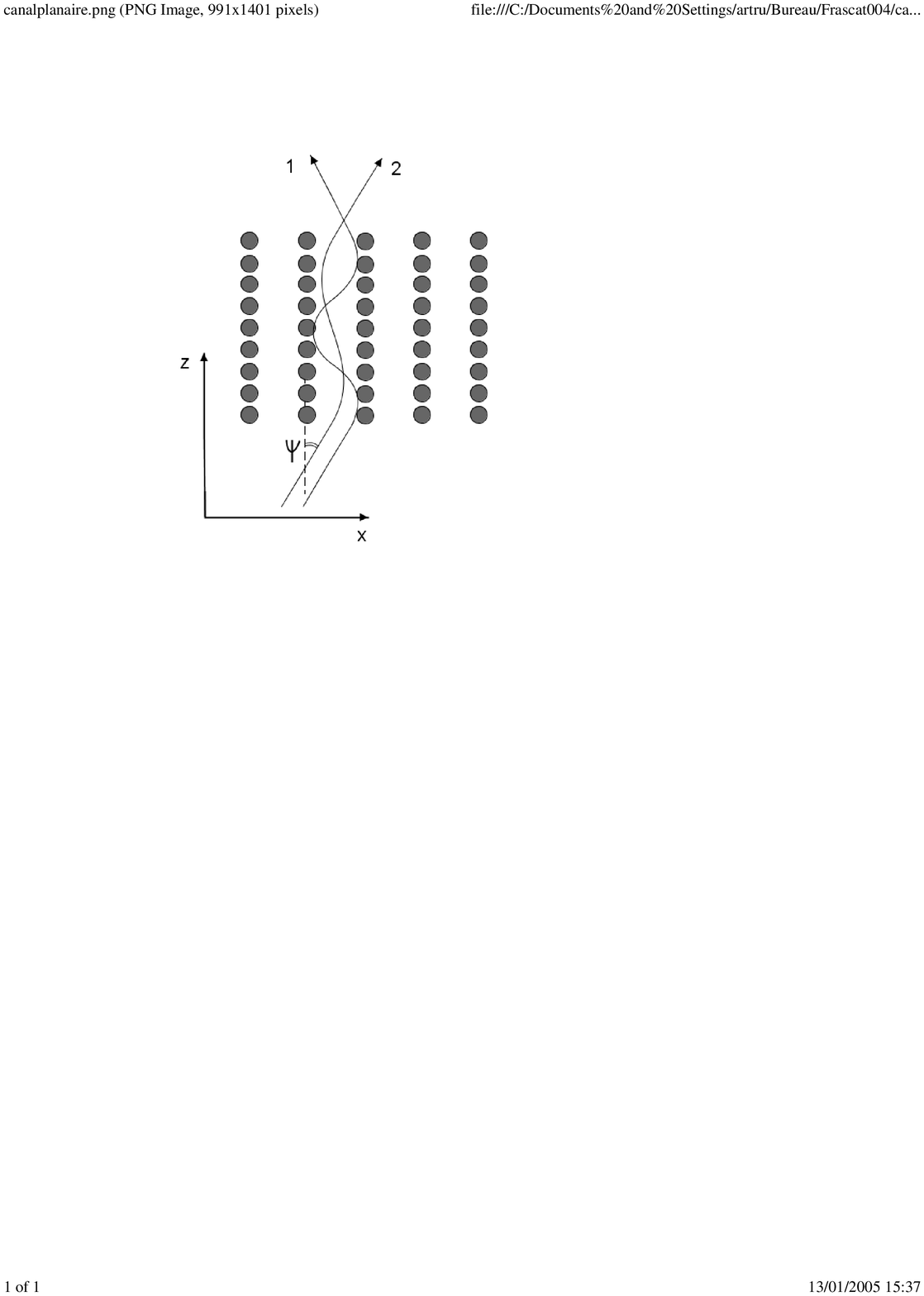}
\caption{two different trajectories of a channeled particle in a crystal, 
for the same entrance angle $\psi$.}
\end{figure}
Neglecting the effect of the non-continuous part of the potential, the parallel momentum 
$\pv_\parallel = (p_y,p_z)$ as well as the so-called {\it transverse energy}
\begin{equation}
E_\perp = U(x) + {p_x^2\over 2 \epsilon}
\,
\label{ET}
\end{equation}
are conserved ($\epsilon$ is the total relativistic energy of the particle). 
The equations of motion are
\begin{equation}
{d\rv \over dt} = {\pv\over \epsilon}
\,,\quad
{dp_x\over dt} = - U'(x) 
\,.
\label{motion}
\end{equation}

Particles with low enough transverse energy are $channeled$, i.e. oscillate 
between two planes if positively charged, about one plane if negatively charged. 
A necessary, but not sufficient, condition for channeling is that the entrance angle $\psi$ 
with respect to the planes be smaller than the critical ($Lindhard$) angle
\begin{equation}
\psi_c = \sqrt {2 U_0 /\epsilon}  
\,.
\label{psic}
\end{equation}
Fig.(1) shows two examples of channeling trajectories followed by positive particles, 
say positrons. They are assumed to be initially parallel (same positive $p_x$), 
however not with the same entrance point, so that their subsequent trajectories are different. 
Positron 1 happens to make an odd number of bounces on the planes and exits the crystal
with roughly reversed $p_x$. 
Positron 2, on the other hand, makes an even number of bounces and exit roughly parallel 
to the incident direction. 
For a beam of many positrons, one expects that the average change 
$ \langle \Delta p_x \rangle $ 
of the electron momentum is opposite in sign to the initial $p_x$.
Due to momentum conservation, the crystal receives an average momentum 
$ - \langle \Delta p_x \rangle $ per electron. 

\bigskip\noindent{\fontsize{11}{15}\selectfont{\bf 
\centerline{3. ENERGY EXCHANGE BETWEEN FIELD} 
\centerline{AND PARTICLE IN A F.E.L.}
}}

\medskip\noindent
Let us briefly recall the principle of a free electron laser.
Consider an helical undulator with the magnetic field $\Bv_u (z)$ 
deriving from the vector potential
\begin{equation}
\Av_u(z) = a_u \ [ \xu \cos (k_u z) + \yu \sin (k_u z) ]
\,.
\label{A-und}
\end{equation}
An ultra-relativistic ($\gamma \equiv \epsilon/m \gg 1$) electron travelling
along the $z$ axis has rotating transverse velocity and momentum given by
\begin{equation}
\epsilon \ \vv_\perp = \pv_\perp = e \Av_u(z) 
\,,
\label{pT-und}
\end{equation}
the longitudinal velocity being
\begin{equation}
\dot z = v_z  
= 1 - { m_{eff}^2 \over 2 \epsilon^2 } = 1 - { 1 \over 2 \gamma_{z}^2 } 
\,,
\label{vz}
\end{equation}
with 
\begin{equation}
m_{eff}^2 = m^2 + \pv_\perp^2 \equiv m^2 (1 + \xi_u^2) 
\,,
\label{meff}
\end{equation}
\begin{equation}
\gamma_z = \gamma \ (1 + \xi_u^2)^{-1/2} 
\,.
\label{gammaz}
\end{equation}
$\xi_u = e \, a_u /m$, often denoted $K$, is the undulator strenght parameter.

Let us assume that a circularly polarized incoming wave $\Ev_{in}(t,z) $,
deriving from the vector potential
\begin{equation}
\Av_{in}(t,z) = a_{in} \ [ \xu \ \cos \omega (t-z) + \yu \ \sin \omega (t-z) ] 
\,,
\label{ain}
\end{equation}
travels in the same direction as the electron. 
This wave transfers energy to the electron at the rate%
\footnote{
In Eqs.(\ref{pT-und}-\ref{force}), the contribution of $ \Av_{in} $ 
to $ \vv_\perp $ and $ \pv_\perp$ is neglected, i.e., we assume that 
$\xi _{in} = e \ a_{in} /m$ is small compared to unity.
}
\begin{equation}
\dot \epsilon = -e \ \vv_\perp \cdot \Ev_{in} 
= - { e^2 \omega \over \epsilon} \ a_u  a_{in} 
\ sin[ \omega t - (\omega + k_u) z ] 
\,.
\label{force}
\end{equation}
%
For most values of $v_z$ this quantity oscillates very fast so that $\epsilon$ 
does not vary significantly. However if $v_z$  is close to the {\it resonant} velocity
\begin{equation}
v_r = {\omega \over \omega + k_u}
\,,
\label{v-resonant}
\end{equation}
the phase of the sine varies slowly and significant energy transfers occur
between the electron and the incoming wave. 
An alternative point of view is that at $v_z \simeq v_r$ the peak frequency 
of the spontaneous undulator radiation in the forward direction is close to $\omega$ 
so that the field radiated by the electron strongly interfere, constructively 
or destructively, with the incoming wave. 

Let us assume $a_u  a_{in} > 0$. 
From (\ref{force}) one can see that an electron close to the resonant velocity
and at $z$ slightly above $v_r t $ gains energy. 
Accordingly, the longitudinal velocity, given by (\ref{vz}), increases.
More generally, this occurs if $z - v_r t - n \lambda$ is slightly positive, 
where $n =$ integer and $\lambda = 2\pi/\omega $ is the incoming wavelength. 
If $z - v_r t $ (more generally, $z - v_r t - n \lambda$) is slightly negative,
the electron loses energy. 
To summarize, the electron is repelled from the hyperplanes 
\begin{equation}
z - v_r t = n \lambda
\,.
\label{hyperplan}
\end{equation}
Similarly, it is attracted toward the hyperplanes
\begin{equation}
z - v_r t = \left(n + {1\over2}\right) \lambda
\,.
\label{hyperplan'}
\end{equation}
In Fig.2 the repelling hyperplanes (dashed lines) follow the diagonals 
of the parallelograms made by two successive vertical lines 
$ z = \Lambda_u \times$ integer and two successive wave front lines 
$ z - t = \lambda \times$ integer ($ \Lambda_u = 2\pi / k_u $ is the undulator period).
The repulsion by hyperplanes is analogous to the repulsion by atomic planes 
for a positive channeled particle in a crystal%
\footnote{
However it would be misleading to view the summits (gray dots) of the parallelograms
as "atoms" of a "space-time crystal".}.
\begin{figure}[ht]
\includegraphics*[scale=0.9,clip,bb=60 380 440 755]{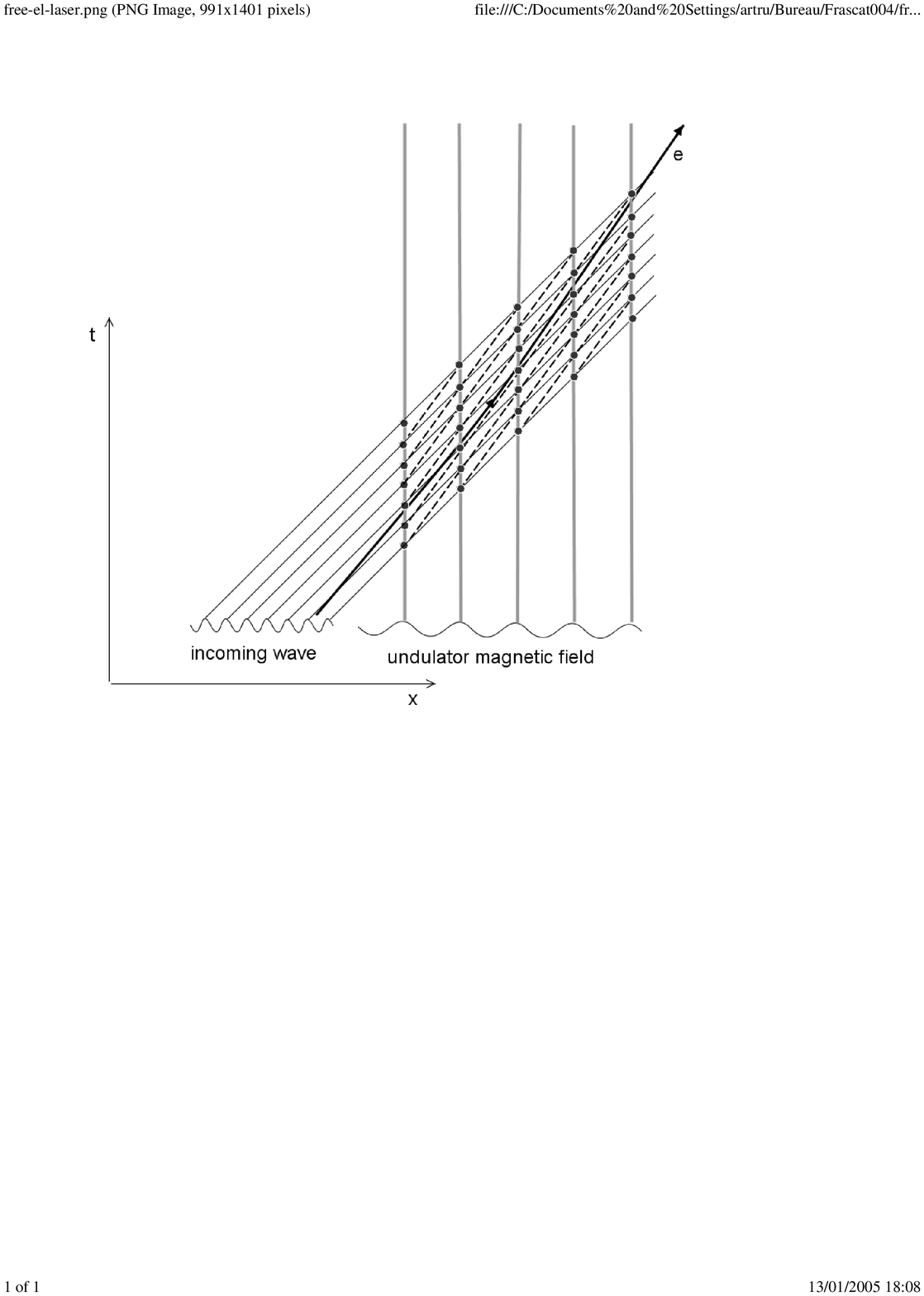}
\caption{world line of an electron or positron in free electron laser}
\end{figure}

Fig.2 shows the trajectory of an electron with initial longitudinal 
velocity%
\footnote{This is the longitudinal velocity just after entering
the undulator field. It differs from the electron velocity before entrance,
due to the effective mass, see (\ref{vz} - \ref{gammaz}). This is a kind of 
refraction in $(z,t)$ space.
}
$v_{z,in}$ slighly above $v_r$.
It bounces once on a repelling hyperplane and exits the undulator with a
longitudinal velocity slightly below $v_r$. 
Due to the fact that $v_z$ is always close to 1,
the bending of the trajectory in the $(z,t)$ plane is not spectacular,
but it corresponds to a significant decrease of the electron energy%
\footnote{
it appears more clearly if one looks at the figure at grazing angle.
}.
This decrease means that the incoming wave has been amplified (stimulated emission).
The value of $\Delta \epsilon = \epsilon_{out} - \epsilon_{in} $ 
depends, for fixed $v_{z,in}$, on the time at which the electron enters the undulator. 
Nevertheless, since $v_z$ oscillates about $v_r$ 
as the electron oscillates between the hyperplanes, 
the average change $ \langle \Delta \epsilon \rangle $ is expected to be negative 
if $v_{z,in}$ is slighly above $v_r$, positive if $v_{z,in}$ is slighly below $v_r$.
This reminds us the correlation between the signs of initial $p_x$ and 
$ \langle \Delta p_x \rangle $ in the channeling case. 
In the case $v_{z,in} > v_r$ the undulator works as a free electron laser.
In the case $v_{z,in} < v_r$ the incoming radiation is partly absorbed 
and the undulator works as an accelerator (inverse FEL).
\begin{figure}[ht]
\includegraphics*[scale=0.9,clip,bb=80 480 455 740]{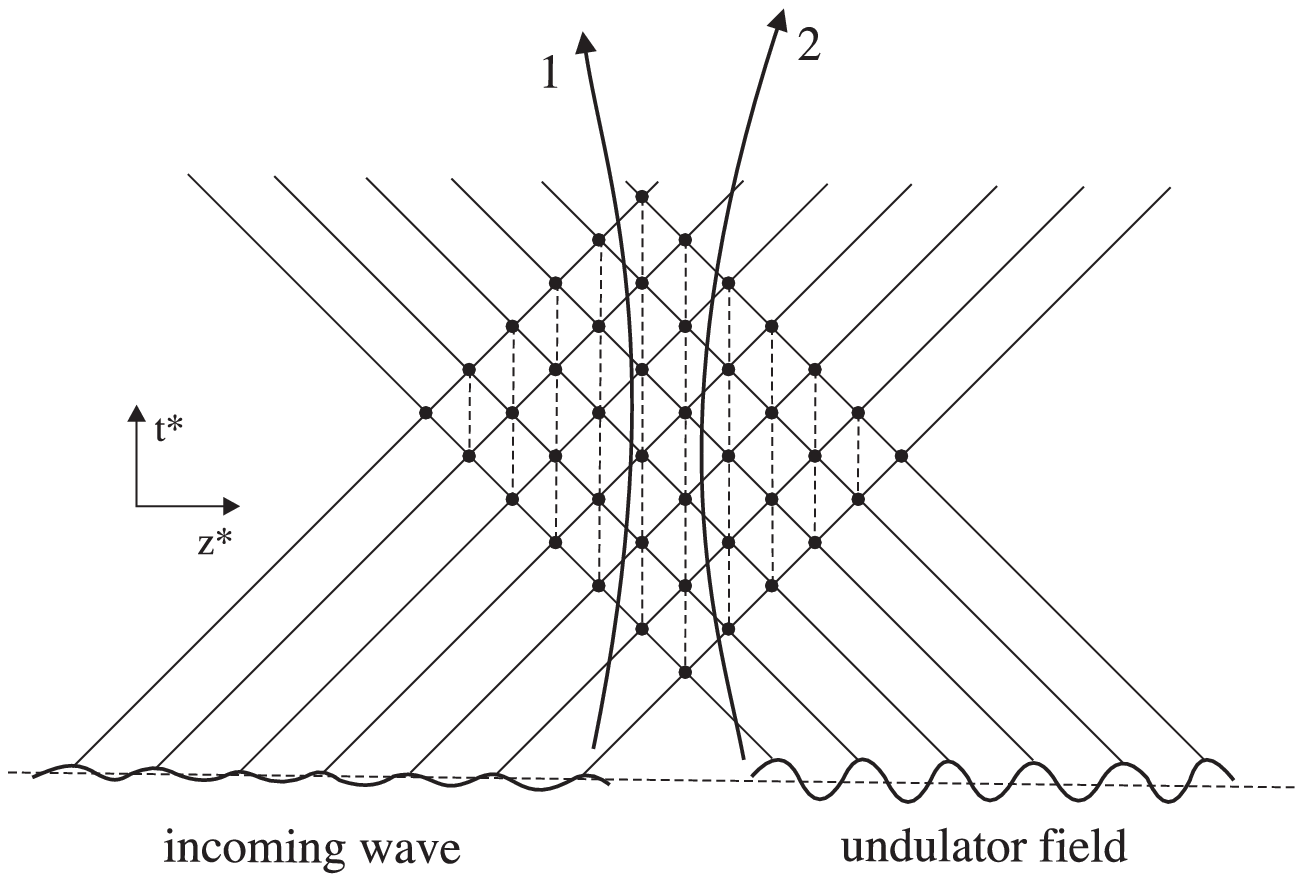}
\caption{electron or positron world line, as in Fig.2, but viewed in the resonant frame.
In the laboratory frame particle 1 is decelerated, particle 2 is accelerated. 
}
\end{figure}

The curvature of the trajectory in longitudinal space-time appears more clearly in the 
{\it resonant frame} $(t^*,z^*)$, which is in translation at the resonant velocity 
with respect to the laboratory frame. In this frame (see Fig.3), the undulator field looks 
like a free, circularly polarized, electromagnetic plane wave of the same angular frequency
$\omega^* = \omega(1+2\omega/k_u)^{-1/2} \sim (\omega k_u/2)^{1/2} $ as the incoming wave. 
The two waves are counter-propagating and build a static {\it ponderomotive potential}
\begin{equation}
U_{pond}(z^*) = { e^2 \over 2 m_{eff}}  \langle \Av^2(t^*,z^*) \rangle
= {\rm constant} + { e^2 \over m_{eff}} a_u a_{in} \cos(2 \omega^* z^*)
\,,
\label{pondero}
\end{equation}
where $ \Av = \Av_{in} + \Av_u $ and $ \langle \ \rangle $ denotes the average over $t^*$.

Similar to the critical channeling angle $\psi_c$, 
one can define a critical velocity in the resonant frame,
\begin{equation}
v^*_c = 2 \ { e \over m_{eff} } ( a_u a_{in} )^{1/2} 
\, 
\label{vcrit}
\end{equation}
above which the particle cannot be trapped in a ponderomotive well. 
The necessary (but not sufficient) trapping condition $|v^*| < v^*_c$ 
writes, in the laboratory frame, 
\begin{equation}
|\gamma_{z}  / \gamma _{z,r} - 1| < v^*_c
\,, 
\label{vcritlab}
\end{equation}
where $\gamma_{z,r}$ is the effective Lorentz factor at resonant velocity, given by
\begin{equation}
\gamma_{z,r} = (1 - v_r^2 )^{-1/2} \simeq \sqrt{\omega /(2 k_u)} 
\,. 
\label{gamma-res}
\end{equation}

\bigskip\noindent{\fontsize{11}{15}\selectfont{\bf 
\centerline{4. CHANNELING IN A BENT CRYSTAL VERSUS}
\centerline{PERIOD-TAPERED UNDULATOR}
}}

\medskip\noindent
If the crystal is bent \cite{channeling}, a channeled positive particle can follow the curvature 
of the planes and then be deflected by an angle much larger than $2 \psi_c$ (Fig.4). 
A necessary (but not sufficient) condition is that the centrifugal force 
$\epsilon v /R$ is smaller than the maximum slope $ U'_{max} $ of the channeling potential:
\begin{equation}
{1\over R} < {1\over R_c} \equiv { U'_{max} \over \epsilon }
\,,
\label{tsyganov}
\end{equation}
$R$ being the curvature radius and $R_c$ is the so-called Tsyganov radius. 
\begin{figure}[ht]
\includegraphics*[scale=0.8,clip,bb=80 545 450 650]{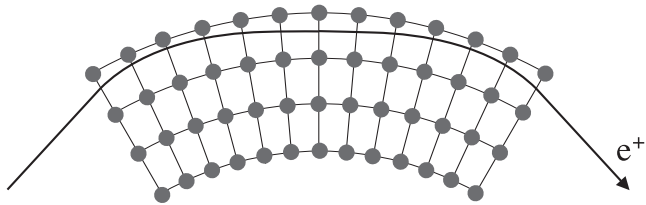}
\caption{deflection of a positron by a bent crystal.
}
\end{figure}

The analog of a bent crystal is a {\it period-tapered} undulator, i.e., an undulator 
where the magnets spacing increases or decreases with $z$. One can see on Fig.5 
that it results in a bending of the "channeling hyperplanes". Hence the particle can suffer
a velocity change in the resonant frame much larger than $ 2 v^*_c $. 
If the period is decreasing with $z$, as in Fig.5, a radiating particle can stay
in resonance in spite of losing energy, since the resonant velocity also decreases.
This phenomenon can be used to extract more energy from the electrons in a FEL
(here we do not consider the other type of tapering which consists in a decrease of 
the amplitude of $|\Bv_u|$ with $z$).
\begin{figure}[ht]
\includegraphics*[scale=0.9,clip,bb=115 420 485 735]{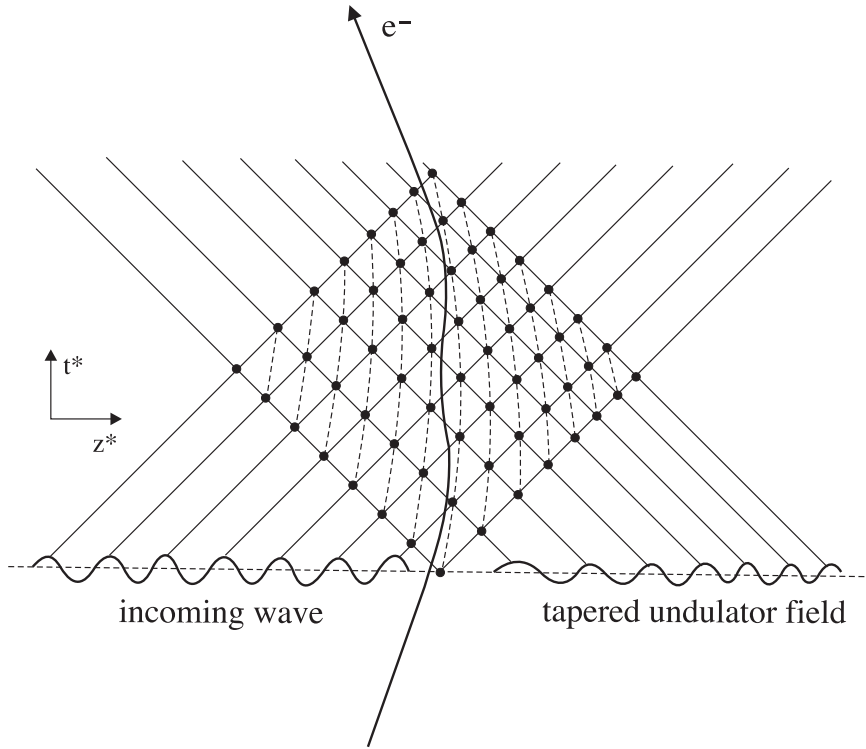}
\caption{trajectory of an electron trapped in the ponderomotive well 
of a tapered free electron laser, viewed in the resonant frame.}
\end{figure}
Conversely, an inverse FEL with an increasing magnet spacing is able to communiquate 
more energy to an electron or positron beam.
The maximum deceleration or acceleration rate (maximum curvature in Fig.5) is proportional 
to the maximum gradient of the ponderomotive potential. In the resonant frame, it is given by
\begin{equation}
\dot v^* _{max} = 
2 \ { e^2 \over m _{eff}^2 } \ a_u a_{in} \ \omega^* 
\,.
\label{gradient}
\end{equation}
This is the analog of Eq.(\ref{tsyganov}). In the undulator frame (see \ref{force}),
\begin{equation}
\dot \epsilon_{max} = m_{eff} \ \dot \gamma_{z} 
= { e^2 \omega \over \epsilon} \ a_u  a_{in} 
\,.
\label{decel}
\end{equation}
Keeping the particle at resonance, i.e. taking $ \gamma_{z}$ = $\gamma_{z,r} $ 
given by (\ref{gamma-res}), we arrive at the maximum tapering \cite{Luchini} 
\begin{equation}
\left|{\Delta \Lambda_u \over \Lambda_u }\right|_{max} 
= 2 \left|{\Delta \epsilon \over \epsilon }\right|_{max} 
= 8 \pi \ {\xi_u \xi_{in} \over 1 + \xi_u^2} 
\,.
\label{tapering}
\end{equation}
where $ \Delta\epsilon$ and $ \Delta \Lambda_u $ are the variations of the electron energy
and of the undulator period $\Lambda_u(z)$ between two consecutive periods
and $\xi_{in}  = e \, a_{in} /m$.

\bigskip\noindent{\fontsize{11}{15}\selectfont \centerline{{\bf 
5. CONCLUSION
}}}

\medskip\noindent
We have seen that the curvature of a channeled particle trajectory in a crystal has a 
particular analogy with the deceleration or acceleration of an electron in a free electron laser.
In the first case the trajectory is curved in the $(x,z)$ plane,
in the second case, it is curved in the $(z,t)$ plane. 
A period-tapered undulator is the analogue of a bent crystal.
To the channeling condition $\psi <\psi_c$ corresponds the trapping condition (\ref{vcritlab})
and to the Tsyganov radius $R_c$ corresponds the maximum tapering rate (\ref{tapering}).

We must say that our presentation of the free electron laser mechanism is far from complete. 
First of all, we have assumed that the electron is trapped inside a well of the
ponderomotive potential, following condition (\ref{vcritlab}). 
In other words we considered only the strong coupling regime.
We also restricted to helical undulators to avoid oscillations of $v_z$ and the generation 
of harmonics. We ignored the interactions between the electrons of a bunch, etc. 
However, we think that the comparison with channeling can help to get a first intuitive 
insight of FEL.

One may pursue the analogy in the weak coupling regime $v^*_c / v^* <1$, 
where the electron jumps from well to well. 
In channeling, it corresponds to {\it over-barrier} particles ($\psi > \psi_c$). 
In the weak coupling, the FEL gain is proportionnal to the derivative of the 
frequency spectrum of the spontaneous radiation (Madey's theorem). It would be interesting 
to see if, after traversal of a straight crystal by a channeled particle, 
the average momentum transfer $ \langle \Delta p_x \rangle $ is given by a similar formula.

\end{document}